\documentclass[10 pt, conference]{IEEEtran}

\usepackage{epsfig,graphicx,color,amsmath,amsfonts,bm,balance,tikz,array,multirow,fancyhdr,float}
\usepackage{url}

\title{\LARGE \bf
Direction of Arrival Estimation for a Vector Sensor Using Deep Neural Networks
}

\author{Jianyuan Yu, William W. Howard,  Daniel Tait and R. Michael Buehrer
\thanks{$*$This work was supported by Wireless @ Virginia Tech}
\thanks{$^{*}$}
\thanks{$^{\dagger}$todo%
       } \\
$^{}$Bradley Department of Electrical and Computer Engineering, Virginia Tech, Blacksburg, VA 24061 \\

\{jianyuan, wwhoward, dtait, buehrer\}@vt.edu
 
}

\begin{document}

\maketitle
\thispagestyle{empty}
\pagestyle{empty}

\begin{abstract}
A vector sensor, a type of sensor array with six collocated antennas to measure all electromagnetic field components of incident waves, has been shown to be advantageous in estimating the angle of arrival and polarization of the incident sources. While angle estimation with machine learning for linear arrays has been well studied, there has not been a similar solution for the vector sensor. In this paper, we propose neural networks to determine the number of the sources and estimate the angle of arrival of each source, based on the covariance matrix extracted from received data. Also, we provide a solution for matching output angles to corresponding sources and examine the error distributions with this method. The results show that neural networks can achieve reasonably accurate estimation with up to 5 sources, especially if the field-of-view is limited.

\end{abstract}


\section{INTRODUCTION}

The vector sensor is a kind of array antenna that consists of two lumped orthogonal triads of scalar sensors that measure the electric and magnetic field components. Its received data is a vector corresponding to the complete electric and magnetic fields rather than the scalar electric field or magnetic field only. Such a sensor can offer greater observability for direction of arrival (DoA) estimation, and requires a smaller array aperture. Earlier work \cite{nehorai1998electromagnetic} introduced the cross product  for the single signal case, and Cramer-Rao lower bound (CRLB) \cite{yuan2011estimating} for the estimation error. Dimensionality Reduction MUltiple SIgnal Classification (DR-MUSIC) \cite{wang2014uni} improves the MUSIC algorithm by reducing the computation  from a single four-dimensional peak search to two sequential two-dimensional peak searches, but neither of them reliably estimates up to 5 sources nor reaches the theoretical limit of five sources. Additionally, the Estimation of Signal Parameters via Rotational Invariance Techniques (ESPRIT) \cite{wong1997uni} requires another time delayed snapshot to  estimate up to 5 sources with single tone modulation.

Moreover, a lot of work has been studied around the application of machine learning on the Uniform Linear Array (ULA) or Uniform Circular Array (UCA), which includes the k-nearest neighbor (KNN) algorithm \cite{roshanaei2009dynamic} and Support-Vector Machine (SVM) \cite{pastorino2005smart}. In particular, \cite{el1997performance} applied neural network to DoA estimation of one signal source using received signal correlation matrix as the input, however, it is not mentioned how to estimate several sources at the same time, and detailed evaluation over low SNR. Another branch of machine learning method, decision tree, is also applied to indoor positioning with Extreme Gradient Boost (XGB) ~\cite{luckner2017application}. Besides the covariance matrix, the input data also can be images ~\cite{adavanne2018direction}, or characteristics of signals like the difference of angles ~\cite{wang2017cifi}, and the knowledge of trajectory ~\cite{abdelbar2017indoor}.

To the end, this paper investigates the behavior of the covariance matrix, and applies them to a neural network to predict multiple DoAs on the vector-sensor. We evaluated several neural networks training options, and applied the sorting method to fix the identification issue and evaluate the estimation error over different angles.


\section{System Model}

The problem is to determine the number of sources and then estimate all directions of arrival, azimuth $\theta$ and elevation $\phi$, based on received signals on 6 antennas.  The auxiliary polarization angle $\gamma$ and polarization phase difference $\eta$ are assumed non-zero constant and are not of interest here.

\subsection{Vector sensor signal model}


\begin{figure*}
  \includegraphics[width=\textwidth,height=4cm]{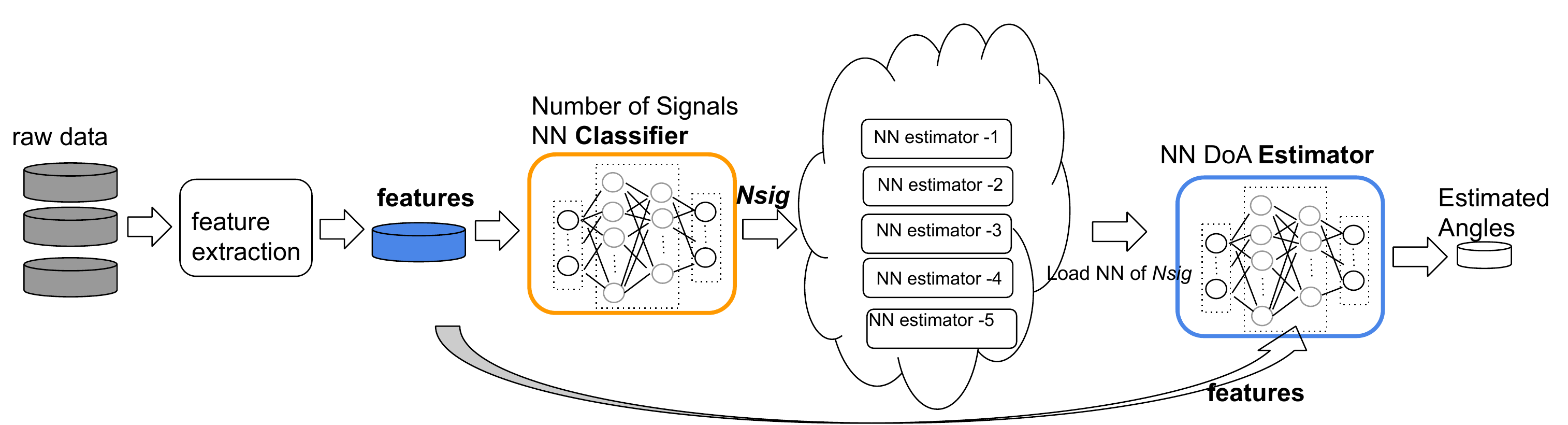}
  \caption{Processing flow of classifier and estimator for overall DoA estimator }
  \label{fig:workflow}
\end{figure*}

The received signal $\textbf{X}_{6 \times N}$ can be described by the model in Wong`s work\cite{wong1997uni}, which is represented as
$ \textbf{X} = \textbf{AS} + \textbf{n} $.
where $\textbf{A}_{6 \times K}$ is the array manifold, $\textbf{S}_{K \times N}$ is the source signal, $\textbf{n}$ is  noise, $K$ is the  number of signal source ranging 1 to 5, and $N$ is the number of snapshots. Meanwhile,   $ \textbf{A }= [\textbf{a}(\theta_1), \textbf{a}(\theta_2),...\textbf{a}(\theta_N)]$, and calculate steering vector $\textbf{a}(\theta)$  by

\begin{equation}
\textbf{a}_k
 =
    \begin{bmatrix}
  \cos \theta_k \cos \phi_k  & -\sin\phi_k \\
  \cos \theta_k \sin \phi_k  & \cos\phi_k \\
  - \sin \theta_k  & 0\\
  - \sin \theta_k  & -\cos \theta_k \cos \phi_k\\
  \cos \phi_k  & -\cos \theta_k \sin \phi_k\\
  0 & \sin \theta_k
\end{bmatrix}
\begin{bmatrix}
  \sin \gamma_k e^{j\eta_k} \\
  \cos \gamma_k
\end{bmatrix} ,
\end{equation}
where $\theta_k \in [0,\pi)$ , and $\phi_k \in [0,2\pi)$. The non-correlated source signal could be single-tone or digital, with any sort of modulation or pulse shaping.







\subsection{Flowchart of the classification and estimation system}
Fig.~\ref{fig:workflow} show how the NN classifier and estimator works. Raw data is converted into feature data by a feature-extraction block then they are fed into a trained NN classifier which estimates the number of sources $K$ present in the data. Finally, the system outputs the angle estimations. Note that the training and test datasets are uniformly distributed over the sphere, instead of over a Cartesian plane. This is because if the data clusters near the poles, the estimation problem becomes more difficult.


\subsection{Data compression by covariance matrix}

\begin{figure}[t]
  \centering
  \includegraphics[width=0.45\textwidth]{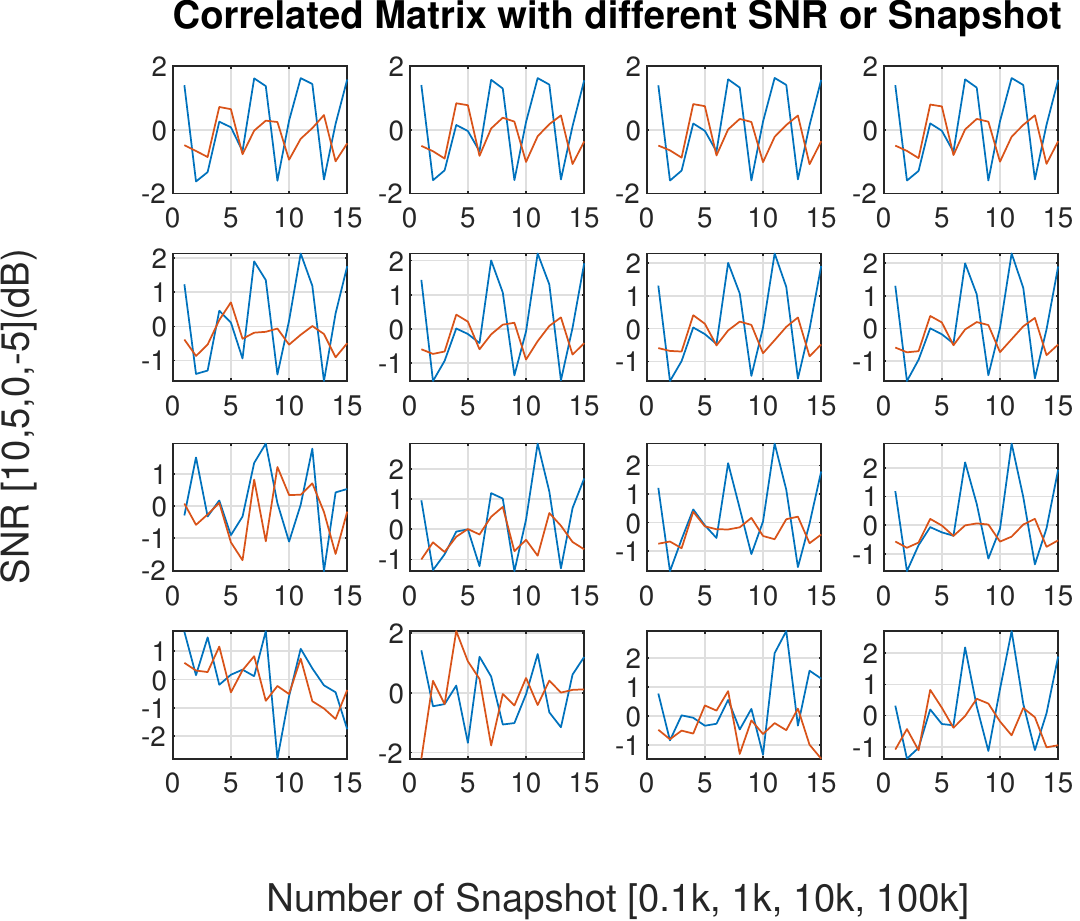}
  \caption{Covariance matrix  keeps its shape at high SNR. Longer snapshots also make the shape more robust to noise}
  \label{fig:CM}
\end{figure}

The covariance matrix (CM) $\textbf{Z}$ is introduced to extract the features related to DoA, as well as a way to compress the data, since the received data $\textbf{X}_{6\times N}$ is too big to directly feed into the neural network.  It is defined as

  \begin{equation}
    \begin{split}
    \textbf{Z} &=\textbf{ X} \textbf{X}^H \\
    &= ( \textbf{AS} + n)( \textbf{AS} + n)^H \\
    &= \textbf{A} (\textbf{SS}^H)\textbf{A}^H + \delta^2 \textbf{I} \\
    &= \textbf{A}\textbf{A}^H\delta_s^2 \textbf{I } + \delta^2\textbf{ I},
    \end{split}
    \end{equation}
\label{equ:cor}where $\delta$ is noise level and $\textbf{A}\textbf{A}^H$ is the only item dependent DoA. We take the right upper triangle of the matrix and split complex into real and imaginary, then convert them into a vector to be fed into NNs later. The covariance matrix can suppress additive white Gaussian noise (AWGN), and remove the dependence of the modulation schemes, pulse shaping or carrier frequency.  It is also adopted as the first step in MUSIC and ESPRIT. The longer snapshot $N$, the more precise the covariance matrix is. A practical rule about the relationship between snapshot length and signal-to-noise ratio (SNR) is that 10 timers longer snapshot length can make up for a 3dB increase in SNR.  Fig.~\ref{fig:CM} demonstrates  CM with a different number of samples and SNR, where the blue line is the real part and the red is imaginary ones. Note that the SNR decreases from high to low for different rows and the number of snapshots increases from left to right for each column. More snapshots are related to a more robust  covariance matrix  at lower SNR. e.g. 100 snapshots would soon degrade from 10dB to 5dB, while $10^6$ snapshots can hold its shape as low as -5dB. The figure also shows the imaginary part also carries valuable information about angles.

\subsection{Fingerprints validation}
\begin{figure}[h]
\centering
    \includegraphics[width=.6\linewidth]{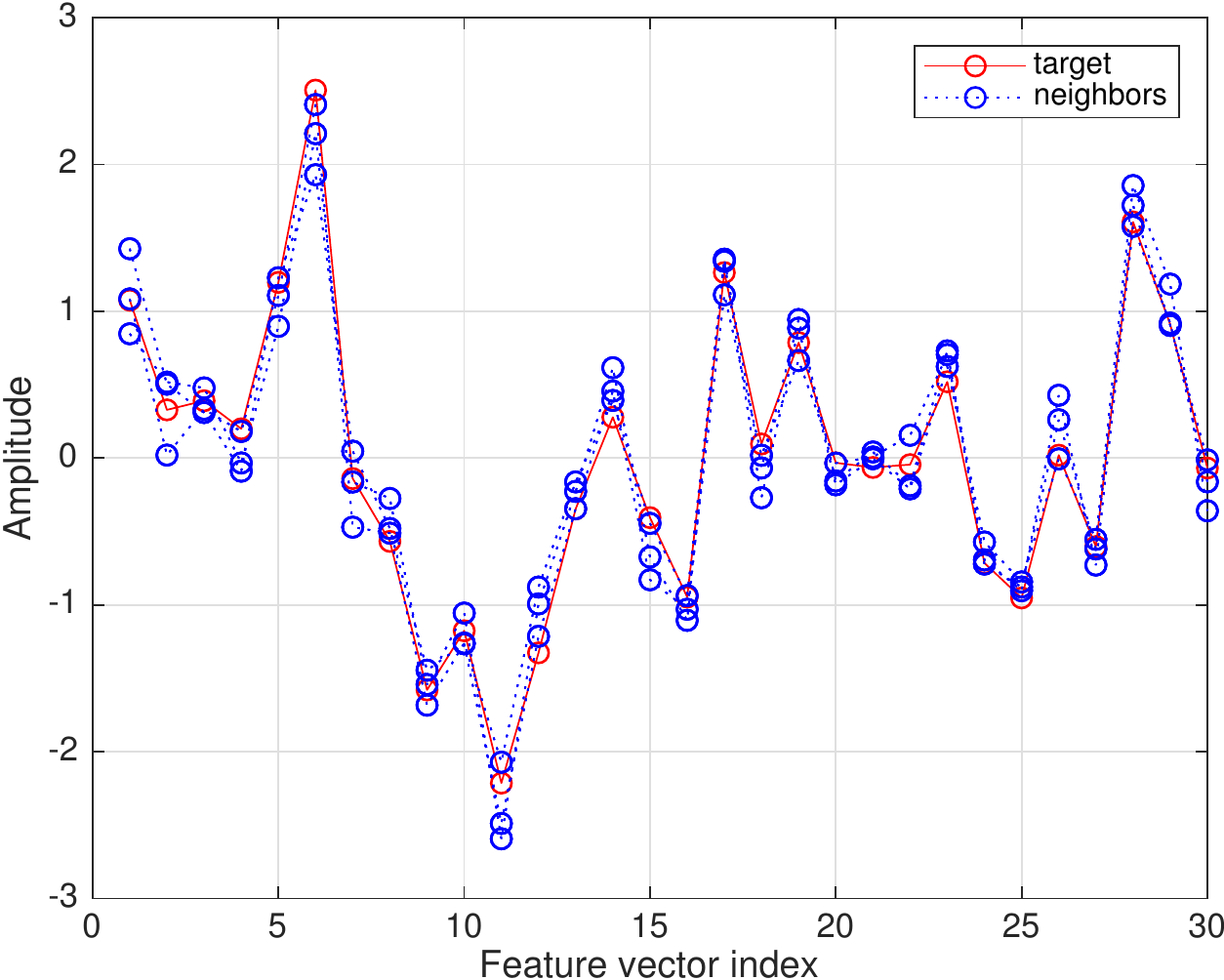}
    \caption{Feature vector of target and its 3 nearest neighbors, Nsig = 5}
    \label{fig:neighbor_vector}
\end{figure}

\begin{figure}[h]
\centering
    \includegraphics[width=0.8\linewidth]{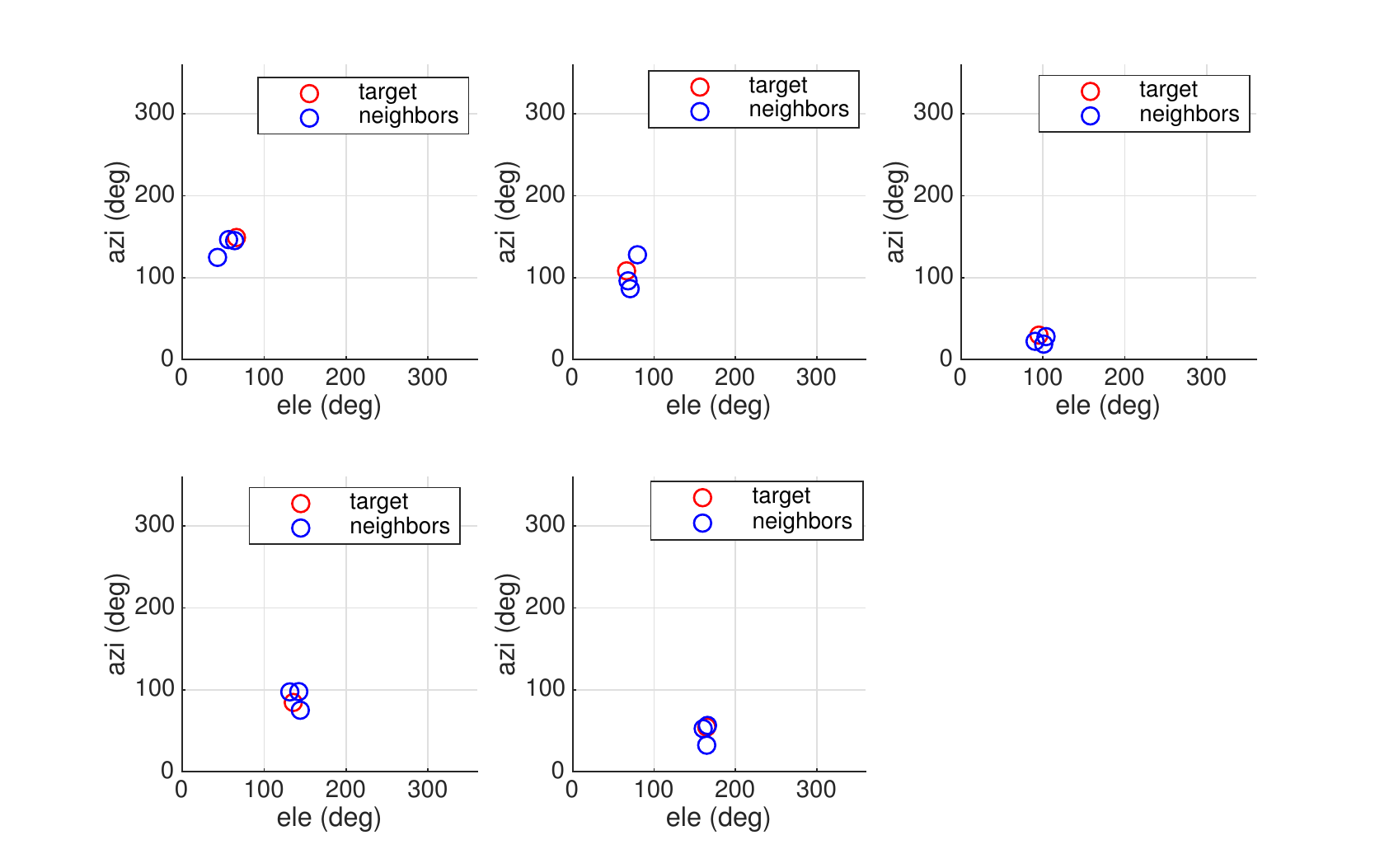}
    \caption{DoA of target and its 3 neighbors, Nsig = 5}
    \label{fig:neighbor}
\end{figure}

The basic idea of all machine learning methods is based on the fingerprint matching system, which is widely used in the DoA estimation and indoor positioning. A basic rule of such a system is the one-to-one mapping property. This property includes two aspects: On the one hand, a similar label should have a similar feature, which is shown in Fig.~\ref{fig:neighbor_vector}. On the other hand, a label has to be associated with a unique feature (namely a \textit{fingerprint}) and the label has to be unique to the feature as well, so that when a new feature is observed we can estimate the label by referring to the neighbors with the most similar features. Fig.~\ref{fig:neighbor} shows that similar features result in similar DoA values allowing the NN to learn values it hasn't seen.




\subsection{Neural network}
From the view of a neural network, the problem is to estimate the DoA of raw data, $ (\hat{\theta_1},\hat{\phi_1}) = f_{NN}(R_{XX}),$ and here $f_{NN}(.)$ is the approximate function given by a neural network. At the training stage, the training data is fed into the neural network and the output provides temporary estimation. The corresponding ground label value is provided to calculate the loss, and the weight of the neural network is updated according to the loss function. The update will repeat many times until the loss reaches a certain minimum. Then at the test stage, the neural network has to predict angles  the data it has never seen.

Several options of the neural network are evaluated here: (i). Levenberg-Marquardt (LM), which has been designed to work specifically with loss functions that take the form of a sum of squared errors. However, LM requires large memory to compute its Jacobian matrix, especially when the datasize is large. (ii). Scaled Conjugate Gradient (SCG): this approach performs the search along with conjugate directions which produce generally faster convergence than regular gradient descent directions. (ii).  Radial Basis Function (RBF),  it claims that by increasing the dimension for easier separation by changing the activation function to a radial form as $\phi(r) = e^{-(\sigma (r-c))^2},$ where $c$ is center vector determined by kNN method. (iv). Bayesian normalization (BNN), to add the weights to overall loss function to restrict the weights from growing too large to prevent overfitting, it does not need validation data but requires a maximum epoch terminate its training.

To better train the NN and prevent overfitting,  several  training techniques are adapted here: (i)  Early-stopping, it splits a validate data set that never trained, but for validation at each epoch, it compares the loss of training and validation each epoch, and stops when these two values diverge. (ii) Drop-out, to avoid a layer relying too much on a few of its inputs, where value grows too large while others are suppressed to zeros. Drop-out method can randomly drop a small part of the input to break the high reliance. (iii) Batch Normalization, it allows each layer of a network to learn by itself a little bit more independently of the other layers. It normalizes the output of a previous activation layer by subtracting the batch mean and dividing by the batch standard deviation.



\subsection{Impact of Field-of-View}

Despite outliers from strong noise, the estimated performance may vary over different angles or sphere regions. We investigated several aspects when field of view affecting the performance. (i). Elevation near the $z$-pole, or DoA facing the $z$-pole: when the object is near the $z$-axis pole with elevation near $0^{\circ}$ or $180^{\circ}$, it is harder to estimate the angles. The reason is that, for the same distance error $d$ on the unit sphere, we have
    \begin{equation}
    \begin{split}
    d & = (r - ~\hat{r})^2 \\
          &= 2(1 - sin(\theta)sin(\hat{\theta})cos(\phi - \hat{\phi}) - cos(\theta) cos(\hat{\theta} ) ) \\
         &= 2(1 - sin(\theta)sin(\theta+ \Delta\theta )cos(\Delta\phi) -  \\
         & \; cos(\theta) cos(\theta+ \Delta\theta ) ).\\
         \end{split}
    \end{equation}
Assuming fixed $\Delta\theta$, the higher the elevation $\theta$, the distance produces a greater error in $\Delta\phi$  meaning larger azimuth error. Notice the elevation only negatively impacts the estimation of the azimuth. (ii). azimuth near $0^{\circ}$ or $360^{\circ}$, or DoA facing positive $x$-pole:  these angles nearby get similar CM features that can confuse the NN. For example, the estimator brings azimuth $359^{\circ}$  given the truth $1^{\circ}$, which only results in  $2^{\circ}$ error in reality, while in evaluation the error can be mistaken as a large one. (iii). azimuth $\phi$ and $\phi+180^{\circ}$: the feature of two angles is similar to each other, despite a few points being the opposite value to each other. (iv). angles near $k\frac{\pi}{4}$: which bring more zero points in the CM. It is significant in one signal while not obvious with multiple sources, in that zero points adding non-zero points result in non-zeros points and ease the problem.
\begin{figure}[h]
  \centering
  \includegraphics[width=0.50\textwidth]{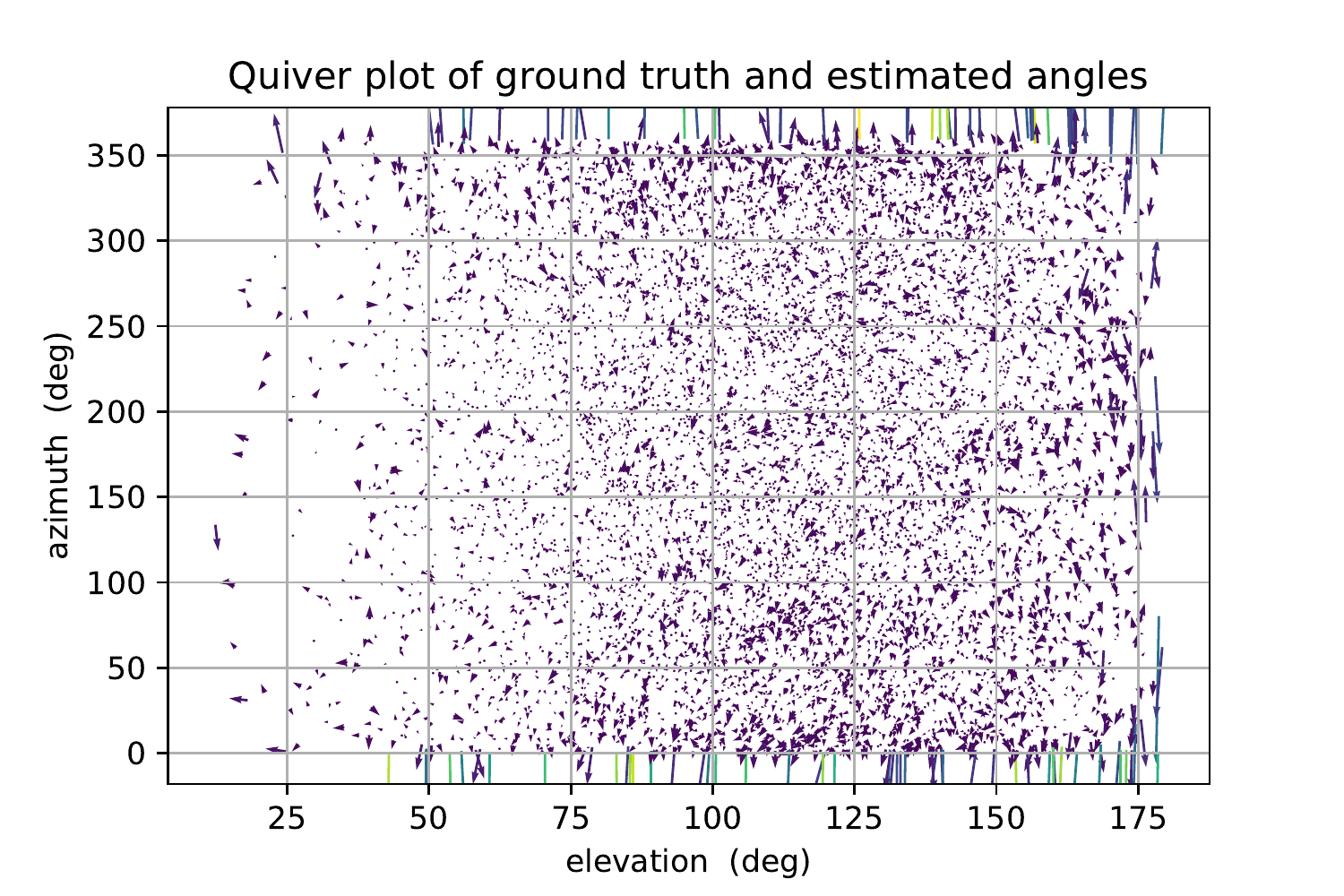}
  \caption{Quiver plot of 2nd DoA estimation error over elevation-azimuth space, Nsig=2 , the head of each arrow means truth and tail means estimation   }
  \label{fig:2_8k}
\end{figure}

The effect of Field-of-view is shown in Fig.~\ref{fig:2_8k}. The estimation performance over the elevation-azimuth space is shown, where the longer and brighter the arrow, the larger estimation error. We found the large error mainly comes from DoAs with azimuth angles near $0^{\circ}$ and $360^{\circ}$, or elevation angles near $0^{\circ}$ and $180^{\circ}$. Notice since we sort the elevation, so the second DoA in the plot has more angles near $180^{\circ}$.






\subsection{Angle identification}
In the case of multiple sources, we only care about the DoA value and the association of azimuth and elevation, but not specific which pair of angles belong to which source. Without decoding the source signal information, there is no way to determine the association based on features. For example, there are 2 sources with angles $(\theta_1= 30^{\circ}, \phi_1 = 60^{\circ}; \theta_2= 20^{\circ}, \phi_2 = 50^{\circ})$, its feature vector is exactly same as that of $(\theta_1= 20^{\circ}, \phi_1 = 50^{\circ}; \theta_2= 30^{\circ}, \phi_2 = 60^{\circ})$. The commutativity would cause a lot of confusion to the NN, especially when calculating the loss function. To this end, we do a simple sort of the elevation on all the dataset , and associated azimuth is sorted simultaneously with elevation. Also the NN will output angle pairs with the elevation  sorted, taking care to sort the test data in the same fashion.

\subsection{Customized loss function }

By default, the loss function is defined as the mean square error (MSE) of angles $ L(\theta,\hat{\theta}) = \frac{1}{N} \sum_1^N{ (\mathbf{\theta}-\hat{ \mathbf{\theta} }  )^2} $, but it does not match the physical meaning of error. In fact, the estimation error should be the distance over the sphere $L(\theta,\hat{\theta}) = \frac{1}{N} \sum_1^N{ (\mathbf{u}-\hat{\mathbf{u}})^2}$,
and  $ \mathbf{u} = (u_x,u_y,u_z)$ by Cartesian-Sphere conversion. We take this metric in defining the loss function of NN.

%

\section{Performance Evaluation}
For the synthesized dataset,  we assume ideal channel conditions where no channel fading, no multiple paths, no interference except white noise, and perfect array geometry without offset. The number of samples is 4000, the signal source is digital, and polarization angles are fixed here.  Azimuth and elevation are both simultaneously randomly generated at float value over the sphere surface. Although the neural network does not require a resolution setting like that in MUSIC, it still expects sufficient data to make interpolation more accurate and error lower. We used MATLAB  to generate a dataset, and Python implements a neural network algorithm. For multiple sources, $512\times5$ hidden layers feed-forward neural network is implemented and trained on a dataset of size $1.6\times10^5$. Means Square Error (MAE) and Root Mean Square Error(RMSE) in degrees are adapted here as the major metrics for estimation error.




\begin{figure}[b]
  \centering
  \includegraphics[width=0.30\textwidth]{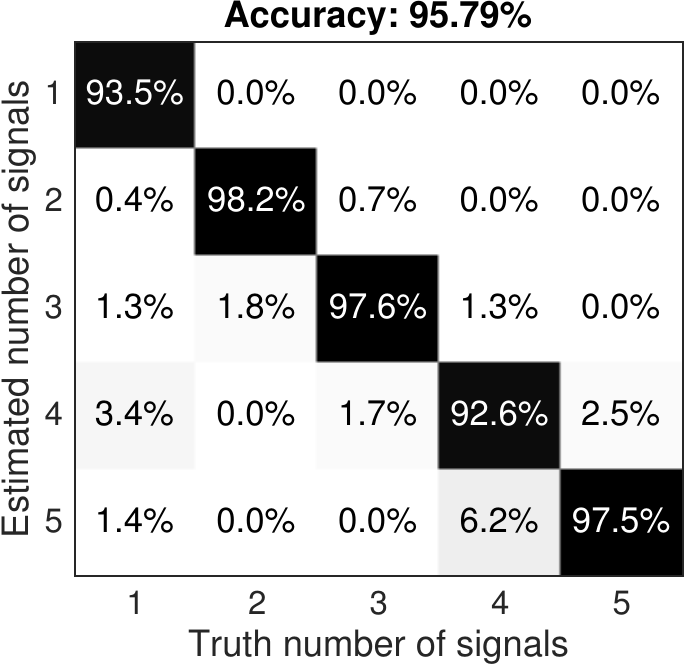}
  \caption{Confusion matrix of NN classifier, random power ratio [-3,3]dB, SNR [0,20]dB}
  \label{fig:conf}
\end{figure}

\subsection{Number of source classification}



The performance of source classification is presented in Fig.~\ref{fig:conf}, it is based on covariance matrix dataset of size $4\times10^6$, SNR in $[0,20]$dB, and random power ratio $[-3,3]$dB related to the first source signal. The classifier can get pretty high accuracy and robust to strong noise. The error mainly comes from distinguishing 4 and 5 sources.



\subsection{Training options }
\begin{figure}[h]
  \centering
  \includegraphics[width=0.5\textwidth]{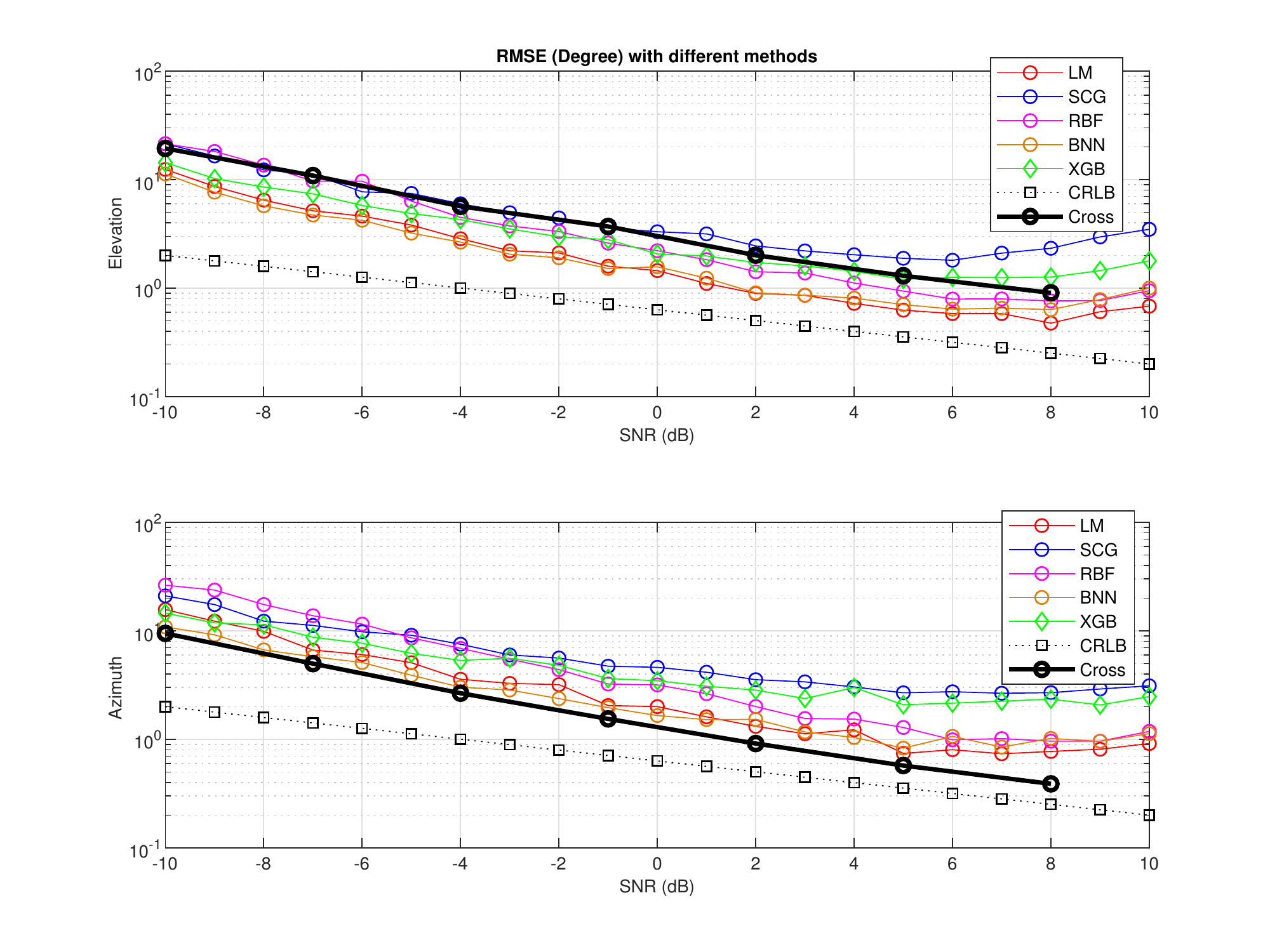}
  \caption{RMSE of different NN-based estimator for one signal }
  \label{fig:SNR1}
\end{figure}

We compare different NN training options here. The BNN maximum epoch is set as 40.  All these models are training with the same data and models are saved down, and then evaluated on the same testing data. The results are shown in Fig.~\ref{fig:SNR1}, with cross-product and CRLB as the baseline.  BNN performs the best and then followed by LM, RBF, and SCG.  All neural-network methods get an advantage over the decision-tree method XGB (extreme gradient boost tree), which accelerates the training speed by gradient descent algorithm. The top three of them can reach 1-degree error after 5dB and close to the cross-product method. Also estimation on elevation is better than azimuth in our settings. Regarding the training time, SCG is fastest with GPU acceleration, and XGB is fastest with CPU only. Hence for larger datasets and more sources case, SCG is preferred afterward.


\subsection{Impact of Field-of-View}

\begin{figure}[h]
  \centering
  \includegraphics[width=0.40\textwidth]{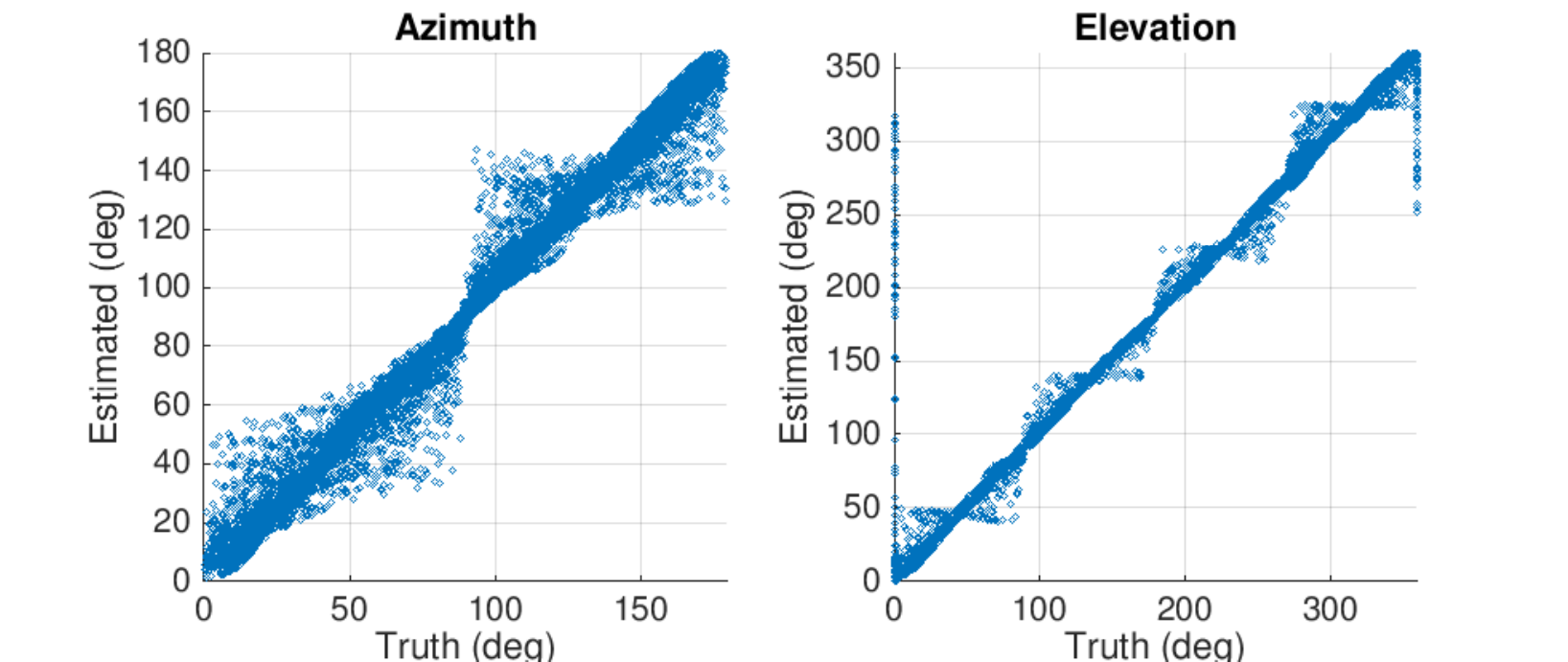}
  \caption{Scatter plot, Nsig =1, no noise, outliners come from $k\frac{\pi}{4}$}
  \label{fig:LimView}
\end{figure}

\begin{figure}[h]
  \centering
  \includegraphics[width=0.35\textwidth]{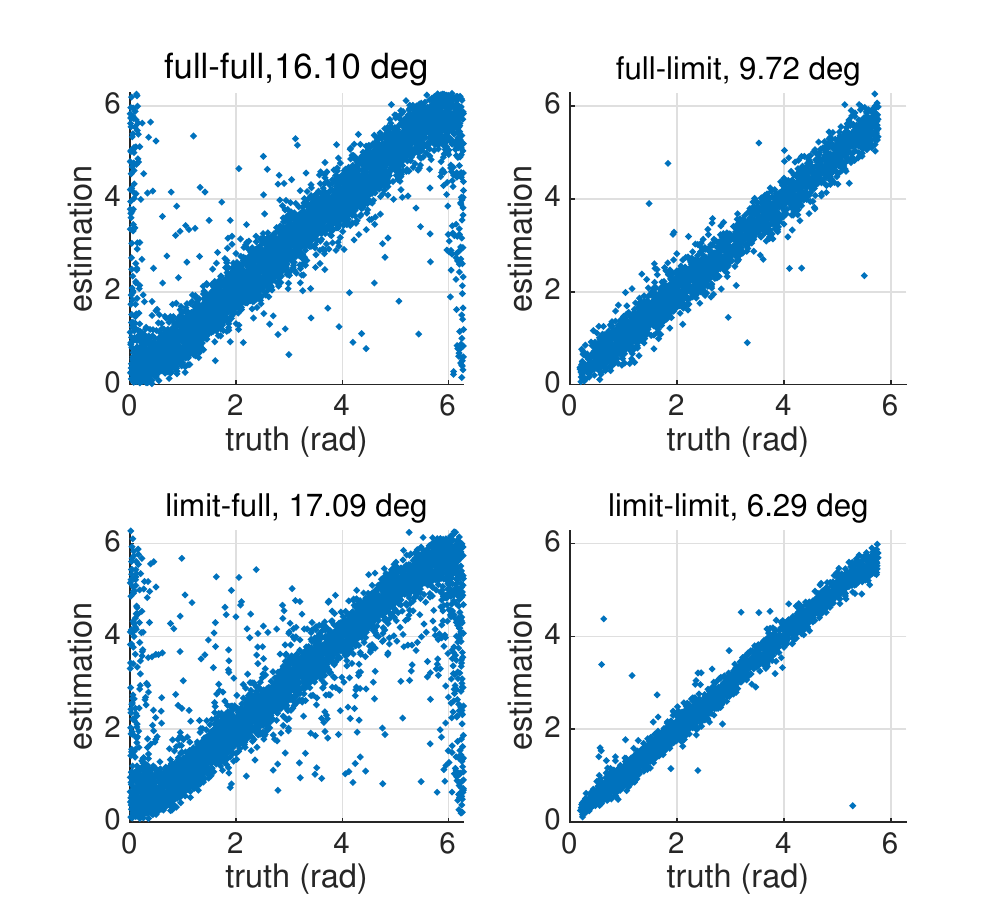}
  \caption{Scatter plot of 1st azimuth estimation, Nsig =5, SNR [0,20]dB}
  \label{fig:cross}
\end{figure}

Fig.~\ref{fig:LimView} shows that the outliers mainly gather at $k\frac{\pi}{4}$ points in the one signal case. While for multiple sources, it mainly degrades when the DoA faces $z$-pole and $x$-pole. In
Fig.~\ref{fig:cross},  word (limited or full) before hyphen mean which data the model is trained on, and word after hyphen mean which data the model is tested on, and here the limited-range we choose azimuth in range $[30^{\circ},330^{\circ}]$ and elevation in $[10^{\circ},170^{\circ}]$. Among these cases, the performance on limit-view test data, based on the model trained on limit-view trained data, is significantly better than others, and we take the limit-view dataset for multiple sources evaluation below.

\subsection{Angles estimation with multiple sources}

\begin{figure}[h]
  \centering
  \includegraphics[width=0.45\textwidth]{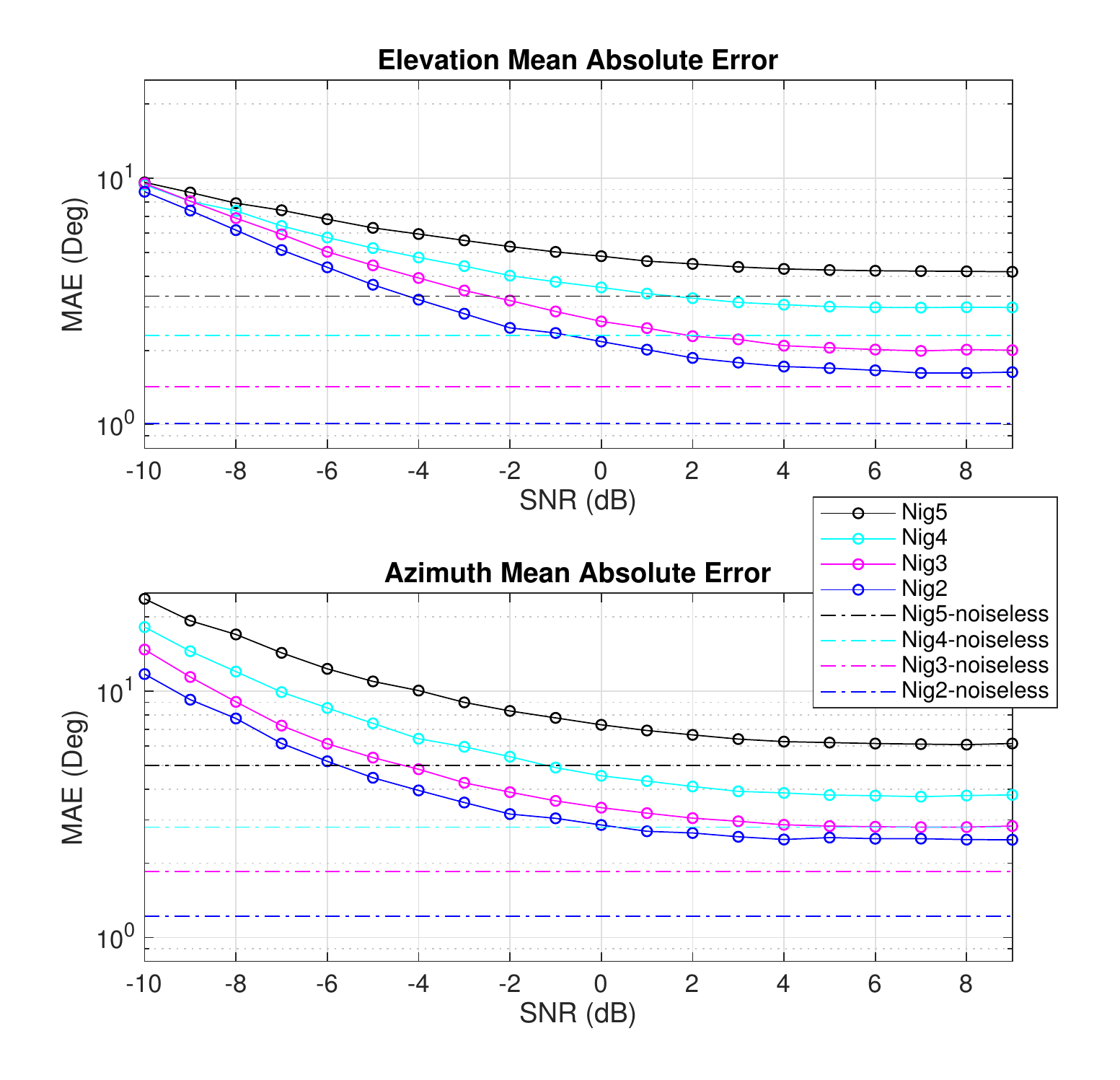}
  \caption{Mean Absolute Error of NN estimator where both trained and tested on data with limited field of view}
  \label{fig:SNR}
\end{figure}

The NN performance of multiple sources is shown in Fig.~\ref{fig:SNR} along with corresponding noiseless cases, which is both trained and tested on the same size data with a limited field of view. We found the more sources, the harder it is for NN to learn to estimate the angles of arrivals and the larger the error. As the SNR goes up, the performance becomes steady after 2dB and comes near its corresponding noiseless case. The NN can reach 5-degree accuracy for both azimuth and elevation with up to 5 sources. Also estimation error of elevation is smaller than that of azimuth.

\section{Conclusions and Further Work}

In this paper, we have shown that deep neural networks can be used to predict the number of sources and the angle of arrival for a vector sensor with satisfactory accuracy. The covariance vector is sufficient to represent the information about angles regardless of the modulation and is robust to noise, while highly reducing the data size needed by the neural network. Additionally, we propose the sorting of input DoAs during training to allow proper identification. Further, we found that the field of view strongly impacts the performance. Our method will be further examined to compare with the ESPRIT and MUSIC methods. Moreover, if the test data keeps its consistency over time, our model could be extended with a recurrent structure to predict an object's trajectory as well.

\addtolength{\textheight}{-12cm}  




\balance
\bibliographystyle{IEEEtran.bst}
\bibliography{IEEEabrv,references.bib}

\end{document}